\documentclass{emulateapj}

\shorttitle{ASTEROSEISMIC DIAGRAMS FROM A SURVEY WITH {\it KEPLER}}
\shortauthors{WHITE ET AL.}
\submitted{Accepted by ApJL}

\begin{document}

\title{Asteroseismic Diagrams from a survey of solar-like oscillations with {\it Kepler}}

\author{
Timothy~R.~White\altaffilmark{1,2}, 
Timothy~R.~Bedding\altaffilmark{1}, 
Dennis~Stello\altaffilmark{1}, 
Thierry~Appourchaux\altaffilmark{3}, 
J\'er\^ome~Ballot\altaffilmark{4,5}, 
Othman~Benomar\altaffilmark{1,3}, 
Alfio~Bonanno\altaffilmark{6}, 
Anne-Marie~Broomhall\altaffilmark{7}, 
Tiago~L.~Campante\altaffilmark{8,9}, 
William~J.~Chaplin\altaffilmark{7}, 
J{\o}rgen~Christensen-Dalsgaard\altaffilmark{9}, 
Enrico~Corsaro\altaffilmark{6}, 
G\"{u}lnur~Do\u{g}an\altaffilmark{9}, 
Yvonne~P.~Elsworth\altaffilmark{7}, 
Stephen~T.~Fletcher\altaffilmark{10}, 
Rafael~A.~Garc\'\i a\altaffilmark{11}, 
Patrick~Gaulme\altaffilmark{3}, 
Rasmus~Handberg\altaffilmark{9}, 
Saskia~Hekker\altaffilmark{7,12}, 
Daniel~Huber\altaffilmark{1}, 
Christoffer~Karoff\altaffilmark{9}, 
Hans~Kjeldsen\altaffilmark{9}, 
Savita~Mathur\altaffilmark{13}, 
Benoit~Mosser\altaffilmark{14}, 
Mario~J.~P.~F.~G.~Monteiro\altaffilmark{8}, 
Clara~R\'egulo\altaffilmark{15,16}, 
David~Salabert\altaffilmark{17}, 
Victor~Silva~Aguirre\altaffilmark{18}, 
Michael~J.~Thompson\altaffilmark{13}, 
Graham~Verner\altaffilmark{7,19}, 
Robert~L.~Morris\altaffilmark{20}, 
Dwight~T.~Sanderfer\altaffilmark{21}, and 
Shawn~E.~Seader\altaffilmark{20} 
}
\affil{\altaffilmark{1}Sydney Institute for Astronomy (SIfA), School of Physics, University of Sydney, NSW 2006, Australia; t.white@physics.usyd.edu.au}
\affil{\altaffilmark{2}Australian Astronomical Observatory, PO Box 296, Epping NSW 1710, Australia}
\affil{\altaffilmark{3}Institut d'Astrophysique Spatiale, UMR8617, Universit\'e Paris XI, Batiment 121, 91405 Orsay Cedex, France}
\affil{\altaffilmark{4}Institut de Recherche en Astrophysique et Plan\'etologie, CNRS, 14 avenue E. Belin, 31400 Toulouse, France}
\affil{\altaffilmark{5}Universit\'e de Toulouse, UPS-OMP, IRAP, 31400 Toulouse, France}
\affil{\altaffilmark{6}INAF Observatorio Astrofisico di Catania, Via S. Sofia 78, 95123, Catania, Italy}
\affil{\altaffilmark{7}School of Physics and Astronomy, University of Birmingham, Birmingham B15 2TT, UK}
\affil{\altaffilmark{8}Centro de Astrof\'{\i}sica and Faculdade de Ci\^encias, Universidade do Porto, Rua das Estrelas, 4150-762 Porto, Portugal}
\affil{\altaffilmark{9}Danish AsteroSeismology Centre (DASC), Department of Physics and Astronomy, Aarhus University, DK-8000 Aarhus C, Denmark}
\affil{\altaffilmark{10}Materials Engineering Research Institute, Faculty of Arts, Computing, Engineering and Sciences, Sheffield Hallam University, Sheffield, S1 1WB, UK}
\affil{\altaffilmark{11}Laboratoire AIM, CEA/DSM-CNRS, Universit\'e Paris 7 Diderot, IRFU/SAp, Centre de Saclay, 91191, Gif-sur-Yvette, France}
\affil{\altaffilmark{12}Astronomical Institute `Anton Pannekoek', University of Amsterdam, Science Park 904, 1098 XH Amsterdam, The Netherlands}
\affil{\altaffilmark{13}High Altitude Observatory, NCAR, P.O. Box 3000, Boulder, CO 80307, USA}
\affil{\altaffilmark{14}LESIA, CNRS, Universit\'e Pierre et Marie Curie, Universit\'e Denis Diderot, Observatoire de Paris, 92195 Meudon cedex, France}
\affil{\altaffilmark{15}Instituto de Astrof\'isica de Canarias, E-38200, La Laguna Tenerife, Spain}
\affil{\altaffilmark{16}Departamento de Astrof\'isica, Universidad de La Laguna, E-38206, La Laguna Tenerife, Spain}
\affil{\altaffilmark{17}Universit\'e de Nice Sophia-Antipolis, CNRS, Observatoire de la C\^ote d'Azur, BP 4229, 06304 Nice Cedex 4, France}
\affil{\altaffilmark{18}Max-Planck-Institut f\"ur Astrophysik, Karl-Schwarzschild-Str.~1, 85748 Garching, Germany}
\affil{\altaffilmark{19}Astronomy Unit, School of Mathematical Sciences, Queen Mary University of London, London E1 4NS, UK}
\affil{\altaffilmark{20}SETI Institute/NASA Ames Research Center, MS 244-30, Moffett Field, CA 94035, USA}
\affil{\altaffilmark{21}NASA Ames Research Center, MS 244-30, Moffett Field, CA 94035, USA}

\begin{abstract}
\noindent Photometric observations made by the NASA {\it Kepler Mission} have led to a dramatic increase in the number
of main-sequence and subgiant stars with detected solar-like oscillations. We present an ensemble asteroseismic analysis
of 76 solar-type stars. Using frequencies determined from the {\it Kepler} time-series photometry, we have measured
three asteroseismic parameters that characterize the oscillations: the large frequency separation ($\Delta\nu$), the
small frequency separation between modes of $l=0$ and $l=2$ ($\delta\nu_{02}$), and the dimensionless offset
($\epsilon$). These measurements allow us to construct asteroseismic diagrams, namely the so-called C-D diagram of
$\delta\nu_{02}$ versus $\Delta\nu$, and the recently re-introduced $\epsilon$ diagram. We compare the {\it Kepler}
results with previously observed solar-type stars and with theoretical models. The positions of stars in these diagrams 
places constraints on their masses and ages. Additionally, we confirm the observational relationship between $\epsilon$ 
and $T_\mathrm{eff}$ that allows for the unambiguous determination of radial order and should help resolve the problem 
of mode identification in F stars.
\end{abstract}


\section{Introduction}
Solar-like oscillations are global modes that are stochastically excited by convection. A revolution
in their study is now taking place thanks to photometric observations being made with NASA's {\it Kepler Mission} 
\citep{Koch10}. Most of the stars are observed at a long cadence of 29.4\,minutes, sufficient for studying
solar-like oscillations in red giants \citep[e.g.,][]{Huber10}. However, up to 512 stars at a time may be observed 
at a short cadence (SC) of 58.85\,s, rapid enough to sample oscillations in main-sequence and subgiant stars 
\citep[e.g.,][]{Chaplin10}. This large number of stars makes it possible to perform large-scale analyses in what has 
been termed `ensemble asteroseismology' \citep{Chaplin11a}. The aim is to correlate observable global
oscillation parameters with physical properties of the stars, such as mass, age and metallicity. This Letter presents a
first attempt at such an analysis for main-sequence and subgiant stars, using a survey of solar-like oscillations made
during the first ten months of the {\it Kepler Mission}. 

\section{Asteroseismic Parameters}

\begin{figure}
\epsscale{1.2}
\plotone{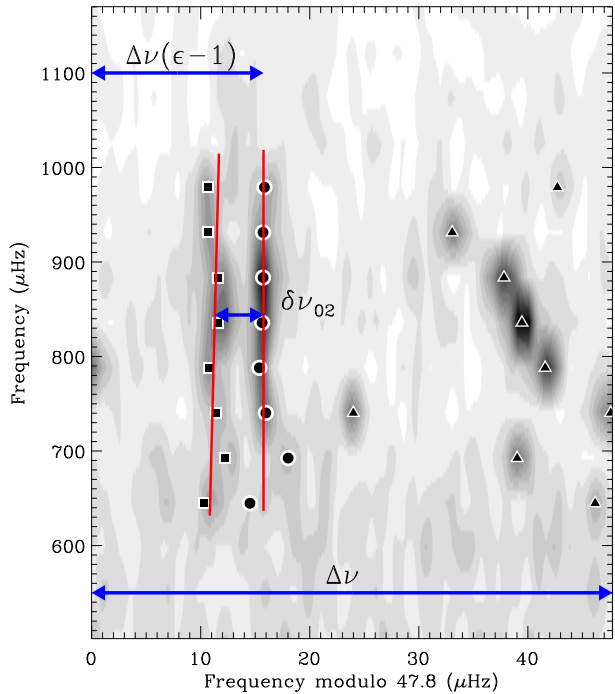}
\caption{\'Echelle diagram of KIC\,11395018 showing the frequencies (black points) as determined by \citet{Mathur11}.
Modes with $l=0$ (circles), $l=1$ (triangles) and $l=2$ (squares) are observed. For reference, a gray-scale map 
showing the power spectrum (smoothed to 1 $\mu$Hz resolution) is included in the background. The fits made to the $l=0$ 
and $l=2$ modes are shown by red lines. The values of $\Delta\nu$ and $\delta\nu_{02}$ (at $\nu_\mathrm{max}$) and the 
relationship of $\epsilon$ to the absolute position of the $l=0$ ridge are indicated by the blue arrows, as labeled.}
\label{fig-1}
\end{figure}

The frequencies of solar-like oscillations for modes of high radial order $n$ and low spherical degree $l$ are well 
approximated by the so-called asymptotic relation \citep{Tassoul80}:
\begin{equation}
\nu_{n,l}\approx\Delta\nu\left(n+{l \over 2}+\epsilon\right)-\delta\nu_{0l}.\label{asymp}
\end{equation}
The large frequency separation, $\Delta\nu$, is the spacing between modes of the same degree and consecutive order and 
is a probe of the mean stellar density. The small frequency separations, $\delta\nu_{0l}$, are sensitive to variations 
in the sound speed gradient near the core in main-sequence stars. This is highly dependent upon the mean molecular 
weight of the core and is therefore a probe of the star's age. The quantities $\Delta\nu$ and $\delta\nu_{02}$ are 
labeled in Figure~\ref{fig-1}, which shows the \'echelle diagram for the {\it Kepler} subgiant KIC\,10395018 
\citep{Mathur11}. 

The different dependencies of the large and small separations have led to the idea of plotting one against the other in
the so-called C-D diagram \citep{C-D84}. Theoretical calculations for the C-D diagram have been carried out for
main-sequence stars \citep[e.g.,][]{Monteiro02,OtiFloranes05,Mazumdar05,Gai09,White11} but prior to {\it Kepler},
observations were only available for a small number of stars. An observational C-D diagram has now been produced for
red giants using {\it Kepler} observations \citep{Bedding10c, Huber10}. In this Letter we present an observational C-D
diagram for main-sequence and subgiant stars.

The third asteroseismic parameter in equation~(\ref{asymp}) is the dimensionless offset $\epsilon$ which, as shown in 
Figure~\ref{fig-1}, is related to the absolute position of the $l=0$ ridge in the \'echelle diagram. The parameter 
$\epsilon$ has been rather neglected until recently, despite also being investigated originally by
\citet{C-D84}. An asteroseismic diagram based on the relation between $\epsilon$ and $\Delta\nu$ was calculated 
from models by \citet{White11} from the zero-age main-sequence to the tip of the red-giant branch. They found that 
evolutionary-model tracks in the $\epsilon$ diagram are more sensitive to the mass and age of subgiants than in 
the C-D diagram. On the observational side, the $\epsilon$ diagram has been published from {\it Kepler} red-giant 
data by \citet{Huber10}. Similar analysis was performed for CoRoT red giants by \citet{Mosser11}.

\begin{figure}
\epsscale{1.2}
\plotone{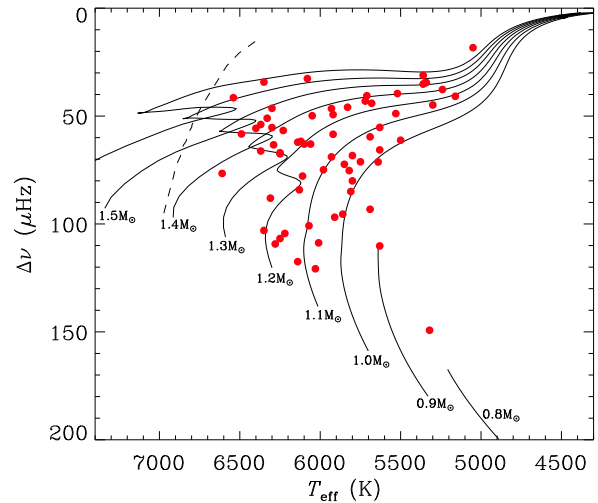}
\caption{A modified H-R Diagram: average large frequency separation, $\Delta\nu$, against effective temperature for
stars in our sample. The lines show solar-metallicity ASTEC \citep{C-D08a} evolutionary tracks for masses between 0.8
and $1.6\,\mathrm{M}_\odot$ in steps of 0.1$\,\mathrm{M}_\odot$. The dashed line indicates the approximate location of the cool edge of
the classical instability strip \citep{Saio98}} \label{fig0}
\end{figure}

\section{Observations and Data Analysis}\label{sec3}

The first ten months of {\it Kepler} SC asteroseismic observations (Q1-Q4) were used for a survey, observing about 2000 
stars for one month each \citep{Chaplin11a}. The time series have been prepared from the raw observations as described 
by \citet{Jenkins10}, and further corrected in the manner described by \citet{Garcia11}. Several methods have been 
developed to detect oscillations and extract global oscillation parameters such as $\Delta\nu$ and $\nu_{\mathrm{max}}$, 
the frequency of maximum power \citep{Campante10,Hekker10,Huber09,Karoff10,Mathur10,Mosser09}. A comparison of these 
methods has been made by \citet{Verner11}. Not all of the observed solar-type stars showed oscillations in the month-long
survey data sets, partially due to the impact of stellar surface activity on the amplitude of oscillations \citep{Chaplin11c}. 

After determining which stars showed oscillations and their approximate values of $\Delta\nu$ and $\nu_{\mathrm{max}}$,
several teams estimated the mode frequencies. In general, each star in our sample was only fitted by a few teams. The
methods for extracting frequencies varied between teams \citep[see][and references therein for details of the different
methods used]{Campante11,Mathur11}. 

Since it is necessary to have clearly identifiable $l=0$ modes to measure $\Delta\nu$ and $\epsilon$, and also to
have $l=2$ modes to measure $\delta\nu_{02}$, we are restricted in our sample to stars that are bright and/or have
the highest amplitudes. Furthermore, it is necessary to have enough identified modes to make a reliable fit. We required
that at least three pairs of $l=0$ and $l=2$ modes were identified in each star. Stars with ambiguous mode
identifications were also not considered. These were generally hotter stars with large linewidths, that made it 
difficult to distinguish between the ridges of $l=0,2$ and $l=1$ modes \citep[e.g., see][]{Appourchaux08,Bedding10}.

To select the final frequency list for each star, the results from different teams were compared with the smoothed
power spectrum. Some teams reported modes for which there was no peak visible in the power spectrum (usually at low or
high frequencies), and those were not included. The remaining lists that passed this comparison could provide
measurements for $\Delta\nu$, $\delta\nu_{02}$ and $\epsilon$ that we deemed reliable. If more than one frequency list
remained for a given star, we chose the one that most closely fitted the observed peaks in the smoothed power spectrum. 
If no reliable fit could be made, the star was excluded from this study. With more {\em Kepler} data, which will become
available for most of these stars, reliable measurements of the separations and $\epsilon$ should be possible in the 
future.  

Our final sample contains 76 main-sequence and subgiant stars with apparent magnitudes ranging between 6.9 and 
11.7. To illustrate their distribution, we show in Figure~\ref{fig0} a modified H-R diagram in which we plot 
$\Delta\nu$ (instead of luminosity) against effective temperature. Temperatures were determined by An et al. (in prep.)
from multicolor photometry available in the Kepler Input Catalog \citep{Brown11}. Solar-metallicity ASTEC evolutionary 
tracks \citep{C-D08a}, neglecting diffusion and core overshoot, are overlaid. Our sample spans an approximate mass range 
of 0.9--1.6 $\mathrm{M}_\odot$. Subgiants dominate the sample because they are intrinsically brighter and their 
oscillations are higher in amplitude.  

From the frequency list for each star we calculated $\Delta\nu$, $\delta\nu_{02}$ and $\epsilon$ using the method
described by \citet{White11}. Briefly, a weighted least-squares fit was made to the radial ($l$=0) frequencies to
simultaneously determine $\Delta\nu$ and $\epsilon$, as given by equation~(\ref{asymp}). The uncertainties in these
parameters were determined from the fit. The fit was weighted by a Gaussian function centered at the frequency of
maximum power, $\nu_{\mathrm{max}}$. This fitting method averages over variation of
$\epsilon$ with frequency (curvature in \'echelle diagrams) that arise from acoustic glitches and the frequency
dependence of the upper turning point \citep{PerezHernandez98,Roxburgh10}. 

In the Sun, $\delta\nu_{02}$ is known to decrease linearly with frequency \citep{Elsworth90}. In calculating 
$\delta\nu_{02}$ from models we performed a linear fit, initially allowing the gradient of $\delta\nu_{02}$ as a 
function of frequency to be a free parameter. Again, the fit was weighted by a Gaussian function and the value of 
$\delta\nu_{02}$ was determined at $\nu_{\mathrm{max}}$. However, for the vast majority of stars, the number and 
precision of the frequencies determined from the {\em Kepler} data do not yet justify the inclusion of the gradient as 
an extra free parameter in the fit. When fitting to the data we have kept this gradient constant at the well-determined 
solar value ($\mathrm{d}\delta\nu_{02}/\mathrm{d}\nu=-0.0022$). In practice, keeping this gradient constant did not 
significantly affect the measured value of $\delta\nu_{02}$, or its uncertainty. 

\section{Results}

\subsection{C-D diagram}

\begin{figure*}
\epsscale{1.07}
\plotone{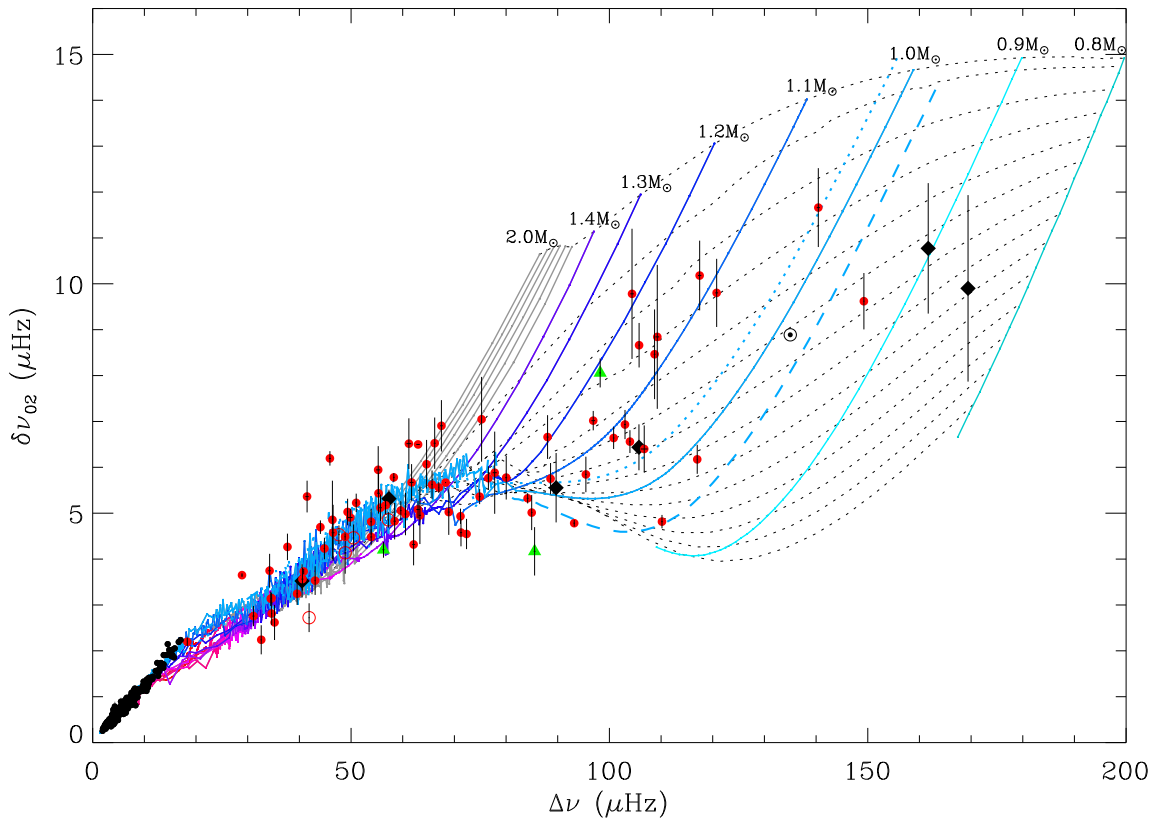}
\caption{C-D diagram, showing $\delta\nu_{02}$ versus $\Delta\nu$. The filled red circles are the 76 {\em Kepler} stars
from this work, while open red circles are {\em Kepler} stars with previously published frequency lists (see text). Also
shown are {\em Kepler} red giants \citep[black circles;][]{Huber10}, and main-sequence and subgiant stars from CoRoT
(green triangles; see text) and ground-based observations (black diamonds; see text). The Sun is marked by its usual
symbol. Error bars show the uncertainties derived from the linear fits for both $\Delta\nu$ and $\delta\nu_{02}$, 
although those in $\Delta\nu$ are generally too small to be visible. Model tracks for near-solar metallicity ($Z_0=0.017$) 
increase in mass by $0.1\,\mathrm{M}_\odot$ from $0.8\,\mathrm{M}_\odot$ to $2.0\,\mathrm{M}_\odot$ (light blue to red lines). 
Also shown are tracks for metal-poor ($Z_0=0.014$; dotted) and metal-rich ($Z_0=0.022$; dashed) solar-mass models. 
The section of the evolutionary tracks in which the models have a higher $T_{\rm eff}$ than the approximate cool edge 
of the classical instability strip \citep{Saio98} are gray: they are not expected to show solar-like oscillations. 
Dotted black lines are isochrones, increasing from 0 Gyr (ZAMS) at the top to 13 Gyr at the bottom.}\label{fig1}  
\end{figure*}

\begin{figure*}
\plotone{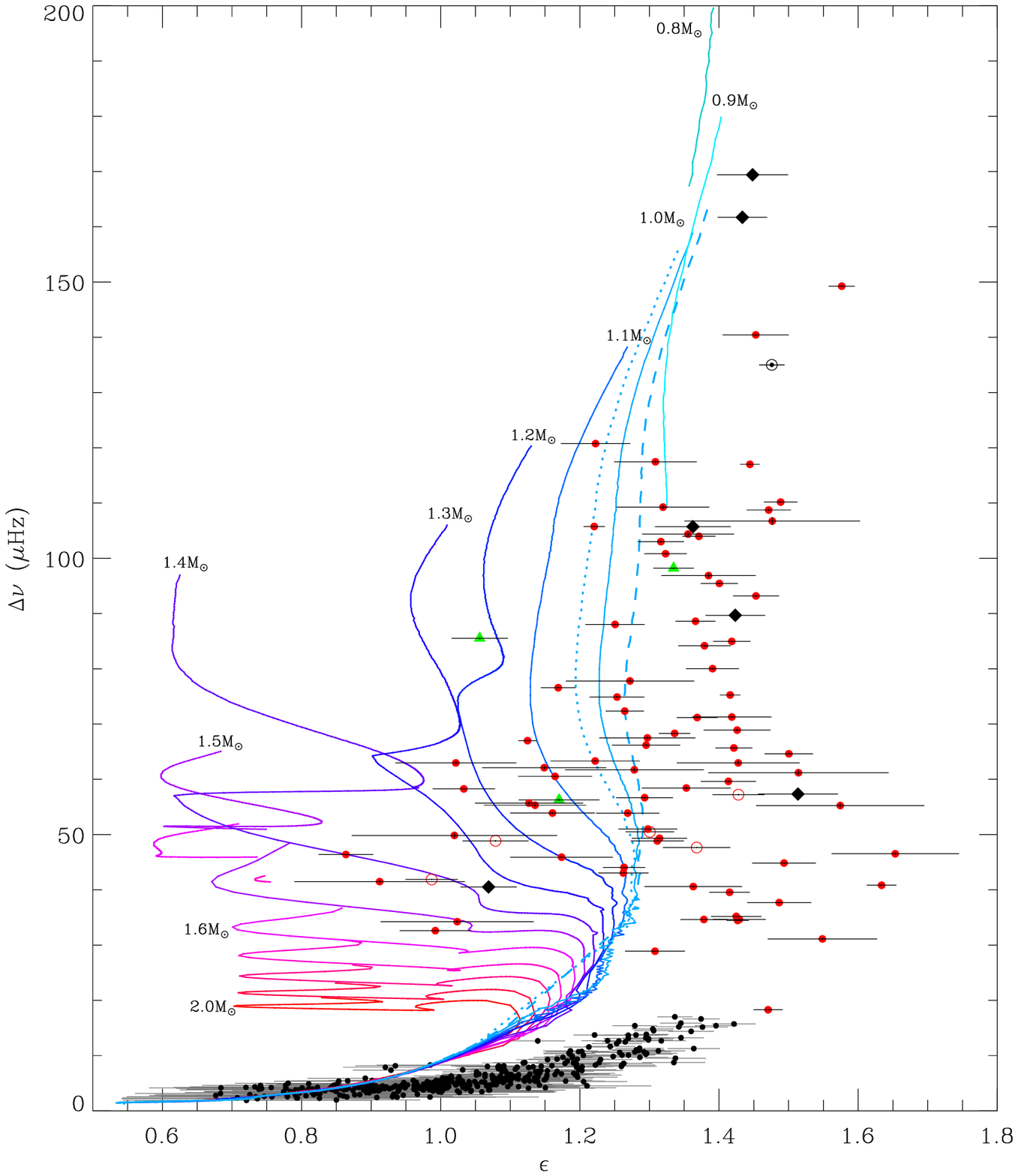}
\caption{The $\epsilon$ diagram, with near-solar metallicity ($Z_0=0.017$) model tracks. Tracks increase in mass by
$0.1\,\mathrm{M}_\odot$ from $0.8\,\mathrm{M}_\odot$ to $2.0\,\mathrm{M}_\odot$. Also shown are tracks for metal-poor ($Z_0=0.014$; dotted) and
metal-rich ($Z_0=0.022$; dashed) solar-mass models. All colors, lines and symbols are the same as for Figure \ref{fig1},
although, for clarity, models in the instability strip are not shown. Error bars show the uncertainties 
from the simultaneous linear fitting of $\Delta\nu$ and $\epsilon$ (the uncertainties in $\Delta\nu$ are mostly too small 
to see).}\label{fig2}
\end{figure*}

Figure~\ref{fig1} shows the C-D diagram for our sample (filled red circles).  
We also include: 
(i) five {\em Kepler} stars with previously published frequency lists (open red circles) --
KIC\,11026764 \citep[known within the Kepler Asteroseismic Science Consortium as `Gemma';][]{Metcalfe10}, KIC\,10273246
(`Mulder') and KIC\,10920273 \citep[`Scully'; both][]{Campante11}, and KIC\,11234888 (`Tigger') and KIC\,11395018
\citep[`Boogie'; both][]{Mathur11}; 
(ii) three CoRoT stars (green triangles) -- HD\,49385 \citep{Deheuvels10}, HD\,49933 \citep{Benomar09}
and HD\,52265 \citep{Ballot11}; 
(iii) six targets observed from the ground (black diamonds) -- $\alpha$~Cen~A \citep{Bedding04},
$\alpha$~Cen~B \citep{Kjeldsen05}, $\beta$~Hyi \citep{Bedding07}, $\eta$~Boo \citep{Kjeldsen03}, $\mu$~Ara
\citep{Bouchy05} and $\tau$~Cet \citep{Teixeira09}; 
(iv) the Sun \citep[usual symbol;][]{Broomhall09}; 
and (v) 470 {\em Kepler} red giants for which $\Delta\nu$, $\delta\nu_{02}$ and $\epsilon$ were measured by 
Huber et al. (2010; black circles).

Stars evolve in the C-D diagram from the top-right to the bottom-left. In Figure~\ref{fig1} we also show the model 
evolutionary tracks as determined by \citet{White11} for near-solar metallicity ($Z_0=0.017$), which are based on the  
ASTEC stellar models shown in Figure~\ref{fig0}. During the main-sequence,
the evolutionary tracks are well separated (at least for masses below $\sim1.5M_{\odot}$) but converge during the
subgiant and red giant phases. Decreasing the metallicity has the effect of shifting the tracks up and to the left, and
in the opposite direction for higher metallicity. We show the tracks for metal-poor ($Z_0=0.014$) and metal-rich
($Z_0=0.022$) solar-mass models for comparison \citep[see][for model C-D diagrams of different metallicities]{White11}. 
To ensure that the observed values are directly comparable to those derived from models, the same recipe for measuring 
$\Delta\nu$, $\epsilon$ and $\delta\nu_{02}$ from data as described in Section \ref{sec3} was also applied to the models.

The positions of the {\it Kepler} stars in the diagram are well covered by the models. In particular, the data show the
convergence of different masses during the subgiant phase, as predicted by the modeling. As a result of this
convergence, the value of $\delta\nu_{02}$ must be measured with high precision for subgiant and red giant stars if it 
is to provide significantly more information about the structure of the star than $\Delta\nu$ alone. This was noted in
the modeling of the {\it Kepler} subgiant KIC 11026764 by \citet{Metcalfe10}. To better constrain the masses and ages of
subgiants, the $\epsilon$ diagram may prove more useful (see Section \ref{eps} below).

The presence of $l\,=\,2$ mixed modes in the oscillation spectrum can complicate the measurement of $\delta\nu_{02}$
by shifting modes from their asymptotically expected frequencies. This is caused by coupling between $p$ modes in the
envelope and $g$ modes in the core \citep{Osaki75, Aizenman77}. The coupling for $l\,=\,2$ is considerably weaker than 
for $l\,=\,1$, but can still 
result in a significant shift in the measured value of $\delta\nu_{02}$. This is one cause of the
scatter for subgiants of the observed values and also of the model tracks, which further increases the difficulty of
determining stellar properties from the values of $\Delta\nu$ and $\delta\nu_{02}$ alone. Metallicity also contributes 
to the observed scatter.

The C-D diagram is primarily useful for main-sequence stars, where the evolutionary tracks are well-separated. If the
position of a star in the C-D diagram can be well determined and the metallicity is accurately estimated, then its mass
and age can be constrained. The physical processes incorporated into the models can also affect the tracks
\citep{Mazumdar05} and so masses and ages determined from the C-D diagram will be model dependent. Measurements of
temperature and luminosity may aid in distinguishing between different models \citep{Monteiro02}.

For the stars in our sample, we do not yet have reliable metallicities, so we are unable to provide
accurate determinations of their masses and ages based upon their position in the C-D diagram. However, we are able to
consider the precision with which these quantities could be determined if we assume a particular uncertainty in
metallicity. We have considered several main-sequence stars of varying mass and age. With an uncertainty in
$[\rm{Fe}/\rm{H}]$ of $\pm0.1$, the masses of these stars could be determined to within 4 to 7\%. Age can generally 
be determined to within 1 Gyr. The precision of the age is heavily influenced by the uncertainty in $\delta\nu_{02}$ 
and the position of the star in the C-D diagram. Many of these stars are being observed for an extended period of time 
with {\it Kepler} and the precision of these measurements will undoubtedly improve. With supporting ground-based 
spectroscopic observations and detailed modeling involving the individual frequencies, the fundamental properties 
of these stars will become well-determined and the data should provide significant tests of stellar models.

\subsection{The $\epsilon$ Diagram}\label{eps}

Figure \ref{fig2} shows the observational $\epsilon$ diagram. Observations cannot determine the radial order $n$ 
directly, and so it is possible for $\epsilon$ to be uncertain by $\pm1$, particularly in the subgiants 
($\Delta\nu$ in the range 20--80 $\mu$Hz). For these, we have taken $\epsilon$ to be in the range 0.7--1.7, 
but note there is some ambiguity for stars near the extremes of this range. The measurement of $\epsilon$ is 
complicated by its close relationship to $\Delta\nu$: a small change in $\Delta\nu$ can induce
a large change in $\epsilon$. We have also measured $\epsilon$ using an alternative method: the variation of
the large separation with frequency was measured as described by \citet{Mosser10}, before the radial modes were globally
fit and $\epsilon$ derived, taking into account the mean curvature. A comparison of the values obtained by the two
methods showed good agreement, although small systematic offsets exist, typically about 0.1. This offset
is probably due to the combined effects of curvature (departure from equation~(\ref{asymp})) and the slightly different 
range of frequencies over which $\Delta\nu$ and $\epsilon$ were measured. A single method must be applied to both models 
and data used to ensure consistency. In this Letter we have used the method outlined in Section \ref{sec3}.

The observed stars in Figure~\ref{fig2} are offset to the right of the
models. This offset is well-known from helioseismology, in which there is a discrepancy between the
observed and computed oscillation frequencies of the Sun arising from improper modeling of the near-surface
layers \citep{Dziembowski88,C-D96}. A rigorous comparison of the observations with models requires
that the offset be taken into account, either by a proper modeling of near-surface layers, or by applying an empirical
correction to the models as suggested by \citet{Kjeldsen08}. It remains to be shown whether this empirical correction is
valid for stars over a wide range of evolutionary states.

As discussed by \citet{White11}, the value of $\epsilon$ may be a significant
asteroseismic constraint on the masses and ages of stars. This is particularly useful for subgiants, for which the
convergence of the evolutionary tracks in the C-D diagram limits the diagnostic potential of the
$\Delta\nu$-$\delta\nu_{02}$ relation. However, it can be difficult in practice to constrain $\epsilon$ because it
varies with frequency, as discussed in detail by \citet{White11}. As apparent in Figure \ref{fig2}, the value of 
$\epsilon$ is sometimes poorly determined. It should also be stressed that the model tracks will vary with metallicity 
\citep{White11}, as can be seen from the solar-mass tracks of different metallicities in Figure \ref{fig2}. 
Nevertheless, for many stars $\epsilon$ is well constrained and with supporting spectroscopy to constrain metallicity, 
the $\epsilon$ diagram shows potential as a useful asteroseismic diagnostic tool.

\begin{figure}
\epsscale{1.2}
\plotone{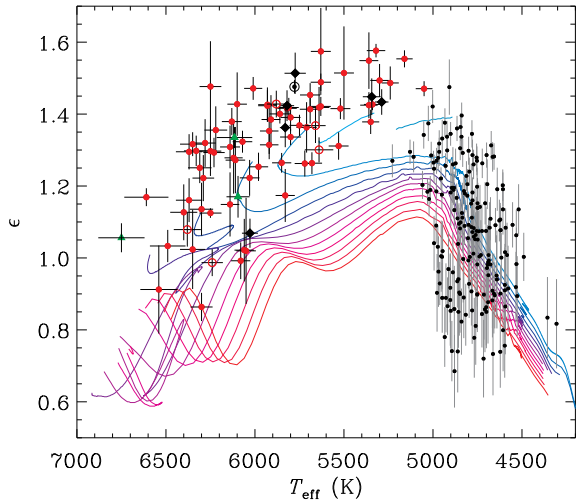}
\caption{$\epsilon$ as a function of effective temperature, with near-solar metallicity ($Z_0=0.017$) model tracks. 
Tracks increase in mass by $0.1\,\mathrm{M}_\odot$ from $0.8\,\mathrm{M}_\odot$ to $2.0\,\mathrm{M}_\odot$. 
All colors, lines and symbols are the same as for Figure \ref{fig1},
although, for clarity, models in the instability strip are not shown.}\label{fig3}
\end{figure}

Finally we note that \citet{White11} demonstrated a relationship between $\epsilon$ and effective temperature. We
confirm this relationship observationally with our larger sample (Figure \ref{fig3}). With this diagram, the 
$\pm1$ ambiguity over the value of $\epsilon$, and therefore radial order, can be resolved. The extension of this 
relationship to higher $T_\mathrm{eff}$ shows promise for resolving the mode identification problem in F stars referred 
to in Section~\ref{sec3} (White et al. in prep).

\acknowledgments
The authors gratefully acknowledge the {\it Kepler} Science Team and all those who have contributed to the 
{\it Kepler Mission} for their tireless efforts which have made these results possible. Funding for the 
{\it Kepler Mission} is provided by NASA's Science Mission Directorate. TRW is supported by an Australian 
Postgraduate Award, a University of Sydney Merit Award, an Australian Astronomical Observatory PhD Scholarship 
and a Denison Merit Award. TRB and DS acknowledge the support of the Australian Research Council. SH acknowledges 
financial support from the Netherlands Organisation for Scientific Research. This research was supported by grant 
AYA2010-17803 from the Spanish National Research Plan. NCAR is supported by the National Science Foundation.

{\it Facilities:} \facility{{\it Kepler}}

\bibliographystyle{apj}
\bibliography{bibliography}

\end{document}